\overfullrule=0pt
\clubpenalty=10000
\widowpenalty=10000

\def\today{\ifcase\month\or
        January\or February\or March\or April\or May\or June\or
        July\or August\or September\or October\or November\or December\fi
        \space\number\day, \number\year}

\def\getcc{\def\endmode{\par\egroup cc: \vtop{\unvbox0}\endgroup\par}\setbox0
       =\vbox\bgroup\hsize=3.5truein\raggedright\leftskip=0pt
\everypar{\hangindent3em}}

\let\endmode=\par
\def\\{\par}


\def\cl{\centerline}

\def\dover#1,#2{\textstyle{{#1 \over #2}}} 

\def\l<{${_<\atop^\sim}$}
\def\r>{${_>\atop^\sim}$}

\def\l{\line}

\def\ni{\noindent}

\def\mm{\itemitem}
\def\vej{\vfil\eject}

\def\ub{\underbar}

\def\ldf{\leaders\hbox to 1 em{\hss.\hss}\hfill}
\def\leaderfill{\leaders\hbox to 1 em{\hss.\hss}\hfill}
\def\dottedline{\leaders\hbox to 1 em{\hss -- \hss}\hfill}

\def\j3m{J.\ Magn.\ Magn.\ Mat.}

\def\boxit#1{\vbox{\hrule height2pt\hbox{\vrule width2pt\kern12pt
        \vbox{\kern12pt#1\kern12pt}\kern12pt\vrule width2pt}\hrule
        height2pt}}

\magnification=1200
\baselineskip=20pt

\cl{\bf Symmetry and Scaling of Turbulent Mixing}
\vskip 50pt

\cl{by}

\cl{by Boris I. Shraiman}
\vskip -9pt
\cl{Bell Laboratories}
\vskip -9pt
\cl{600 Mountain Avenue}
\vskip -9pt
\cl{Murray Hill, NJ  07974}

\cl{\&}

\cl{Eric D. Siggia}
\vskip -9pt
\cl{Cornell University}
\vskip -9pt
\cl{Laboratory of Atomic \& Solid State Physics}
\vskip -9pt
\cl{Ithaca, NY  14853-2501}
\vskip 50pt

\cl{\ub{Abstract}}

The stationary condition (Hopf equation) for the ($n$+1) point
correlation function of a passive scalar advected by turbulent
flow is argued to
have an approximate $SL(n, R)$ symmetry which provides a starting point for
the perturbative treatment of
less symmetric terms.   The large scale anisotropy is found to be a
relevant field, in contradiction with Kolmogorov
phenomenology, but in agreement with the large scalar skewness
observed in shear flows.  Exponents are not universal, yet
quantitative predictions for
experiments to test the $SL(n, R)$ symmetry can be formulated in terms of
the correlation functions.
\vej

Kolmogorov succeeded, by very simple arguments (K41),  in predicting the scaling
for the velocity
correlations $<\upsilon_k \upsilon_{-k}>$ at high Reynolds number $R$, for
wavenumbers intermediate between
those defined by the geometry of the flow and dissipation; the so called
inertial range.$^1$  Obukhov and Corrsin soon
applied the K41 reasoning to a passive scalar i.e., a field $\Theta$ obeying,

$$\partial_t\Theta + \vec\upsilon \cdot \vec\nabla\Theta =
\kappa\nabla^2\Theta\eqno(1)$$

\ni
($\kappa$ is diffusivity), and experiments subsequently found
scaling behavior for $\Theta$, although the K41 exponent is approached
only at very high R if at all.$^2$  However an even more glaring
inconsistency with K41
appeared with the observation that the derivative skewness
$s_d=<(\partial_x\theta)^3>/<(\partial_x\theta)^2>^{3 \over 2}$ is of order
one and $R$ independent out to highest $R$ available.$^3$
This fact motivated our work.

Since $s_d$ breaks parity we follow the experiments and
impose a large scale gradient $\vec g$ so that $\Theta = \theta + g \cdot
r$ where $<\theta> = 0$ and
$<\theta(r)\theta(0)>$ inherits a correlation
length or integral scale from the velocity field.  The inhomgenous term
$\vec g \cdot \vec v$ which appears
in the $\theta$ equation then acts as a ``force" which maintains the
$\theta$ fluctuations stationary.  The
conundrum with $s_d$ is that K41 predicts small scale isotropy for large
$R$ specifically, $s_d \sim g/<(\partial_x\theta)^2>^{1 \over 2}
\sim R^{- {1 \over
2}}$ or for $\delta\theta_r =
\theta(r) - \theta(0)$, $S = <\delta\theta_r^3> \sim r^{5 \over 3}$, (vs
$r^1$ in experiments).  The K41 arguments have proved so seductive and work
so well
for the 2-point velocity correlations that
problems with the scalar and difficulties with higher
order velocity derivative statistics in simulations
have been all but forgotten.

The observation of $s_d \sim O(1)$ is particularly intriguing since it
suggests that $\vec g$ is a $relevant$ variable along with the ``energy"
(scalar variance really) dissipation rate $\epsilon$, in the sense of
parametrizing the effect of the large,
geometry specific scales on the inertial range.
Averaging $\vec g$ away by insisting on isotropic large scales
will eliminate $s_d$, but in no
way lessen the instantaneous non local interactions between large and
inertial range scales.  Whether this
merely obscures the relevant physics, much like studying superfluid Helium
in a fixed number ensemble, or
actually eliminates the dominant exponent for even order correlation
functions remains to be seen.

The prospects for an analytic theory of scalar intermittency are much
enhanced by the observation that $s_d$
remains $\sim 1$ and $R$ independent even when the complex turbulent
shear flow is replaced by a Gaussian
$\upsilon$ field.$^4$  Other simulations and a plausible
closure scheme of Kraichnan
make the same point for even powers of $\delta\theta_r$$^5$: boring but scale
invariant velocity fields generate
interesting scalar statistics.

In this report we argue that the dominant term in the evolution
equation (the ``Hopf equation") of the scalar multipoint correlator
$\psi$ generated by a velocity field with a
physical correlation time is highly symmetric and is integrable by Lie Algebraic
methods.
The leading (anomalous scaling) terms in $\psi$ arise as the
zero modes of the Hopf operator,
$L_H$, with the lowest scaling index $\lambda$.$^{6,7,8}$
For the dominant, symmetric part of $L_H$, the zero modes are labeled by a
complete set of quantum numbers, but because of the high symmetry the
$\lambda$ are infinitely degenerate, i.e., independent of a subset
$\lbrace q \rbrace$ of the quantum numbers.
Smaller, lower symmetry, perturbations, which are nevertheless
dominant  when points are collinear
are treated
by singular perturbation theory, or equivalently by diagonalizing within
the degenerate subspace.
Like the operator itself, the exponents and correlator depend
on the details of the velocity field beyond its scaling dimension.
However since $\psi$ with a general configuration of points
is governed approximately by the symmetric part of $L_H$, it
can be represented as a linear superposition of the degenerate
$\lbrace q \rbrace$ modes with the weight defined by $\psi$ with
points collinear.  The latter can be measured directly in
experiment or numerical simulations and the above relation used
as a quantitative check for the existence of the $SL(n,R)$ symmetry.

The homogeneous part of the inertial range Hopf equation for a white noise
velocity field scaling as $<\delta
\upsilon^2_r> \sim r^\zeta$ may be written in terms of the Richardson operator

$$L^{(\zeta)}_R = r_{ij}^{2-\zeta}\Bigl(\delta^{ab}(d + 1 - \zeta) - (2 -
\zeta)\hat r_{ij}^a \hat
r_{ij}^b\Bigl) \partial^a_i\partial^b_j\eqno(2)$$

\ni
acting on all pairs of points $(i, j, \vec r_{ij} = \vec r_i - \vec r_j)$
and where $a, b$ denote spacial
indices.$^9$  However within K41 theory $(\zeta = {2 \over 3})$ the velocity is
not white but has a scale dependent
(Lagrangian) correlation time $\sim r^{2 \over 3}\epsilon^{-1 \over 3}$.
This is just a restatement of the
scale invariance of the velocity equation which says that the change in
spacing $r$ of two points over the
correlation time of their relative velocity,
$\delta\upsilon_r$, is $\sim r$.  Take the spacial arguments of some
multipoint correlation roughly
equidistant, let $R_g$ be of order their radius of gyration, then a first
approximation to the action of the velocity
field is an $O(1)$ volume preserving linear mapping of all points.
For Gaussian $\upsilon$, the averaged operator
implementing this finite coordinate change is just the exponential of $L_0 =
L^{(2)}_R$, so the condition of
stationarity can be expressed as $L_0\psi = 0$.  (Note $L_0$ is
identical with the Batchelor-Kraichnan
operator$^{10}$ for white random gradient advection but its meaning here is
quite
different.) The action
of $\upsilon$ from scales less than $R_g$ is
incoherent and uncorrelated between points
and plausibly white in comparison with the coherent part. Upon averaging
it gives rise to an ``eddy dissipation" operator $L_D$ (acting along with
$L_0$) the form of which we will discuss later. Finally, the velocity on
scales $\gg R_g$ is an overall
translation which in the presence of the large scale gradient generates
an inhomogeneous forcing term, which however does not affect the zero
modes.

The operator $L_0$ is a very attractive starting point for perturbation
theory since it is integrable.  For the $(n +1)$ point correlation
function of $r_i$ eliminate the center of mass by defining
$n$ difference vectors, $\rho_1 = (r_1 - r_2)/\sqrt 2$, $\rho_2 = (r_1 +
r_2 - 2r_3)/\sqrt 6$, $\rho_3 = (r_1
+ r_2 + r_3 - 3r_4)/\sqrt {12}$, etc.

$$L_0(n) = -(d + 1) L^2 + 2dG^2 + d (d - n) \Bigl(\Lambda^2/(n d) +
\Lambda\Bigr)\eqno(3)$$

\ni
where $L^2 = {-1 \over 2} {\sum \atop {a, b}} (\rho^a_i\partial^b_i -
\rho^b_i\partial^a_i)^2$ is the
square of the total angular momentum; $\Lambda = \rho^a_i\partial^a_i$, the
dilatation operator; and
$G^2 = {1 \over 2}\sum G_{ij}G_{ji}, \ G_{ij} = \rho^a_i\partial^a_j - {1
\over n} \delta_{ij}\Lambda$, is
the Casimir of the group $SL(n, R)$ acting on the $i -$ label, diagonal in
space, henceforth called isospin. The appearance of the $SL(n,R)$ in (3) is
the consequence of the underlying evolution step being the
multiplication of all $\vec \rho_i$ by a spatial strain matrix---the linear
mapping described earlier---which is invariant with respect to a
general linear transformation acting on $\rho$'s.
One might have expected the group $SL(d, R)$ (Casimir $J^2$) to appear,
since we are applying unimodular
matrices in space, and it does if we exploit a ``duality" relation on $G^2$
to replace the last 3 terms in $L_0$ by $2dJ^2$.

The zero modes of $L_0$ are found by constructing the
representations of $SL(n) \times
SO(d) \times {\Lambda}$.  The $SL$ groups are non compact, their
representations infinite dimensional,
and those of interest non unitary---we do not have a suitable inner product.
We work in the space of
homogeneous functions whose scaling dimension, $\lambda$, diagonalizes
$\Lambda$.  The remaining quantum
numbers are discrete.  As before we define $\nu$, related to $\lambda/2$ by
an integer, (defined below)
in terms of which $G^2 \rightarrow (n-1)(2\nu/n+1)\nu$ (for $n\leq d$).
These quantum numbers,
together with those for $SO(n)$
(which is one choice for the maximal commuting sub algebra of $SL(n)$) and
$SO(d)$ are a maximal set for
$(n, d) = (2, 2)$ and the corresponding eigenfunctions were given in [9].
The most relevant (i.e., smallest
non negative[11]) eigenvalues $\lambda$ for any $(n, d)$ are $\lambda = 1$ for
$p$ - wave, $l = 1$; and $\lambda
= 0$ for $s$ - wave.  The latter could have been anticipated from the
Batchelor-Kraichnan interpretation of
(3).  Finally we note that acting on any function behaving as
$|r_{ij}|^x$, $x > 0$ the operator $L_0(n)$ reduces in the limit $r_{ij}
\rightarrow 0$
smoothly to $L_0(n-1)$ acting on the remaining variables. This is because
$L_0$ originates from the advective part of
(1) which evolves $\theta ^2 (r)$ the same way as $\theta (r)$, allowing
the different points in the correlator to be brought together with
impunity.

The dissipation in (1), and its restatement in terms of $L_D$, the
incoherent eddy damping part of the Hopf operator,
of course can not reduce so simply.  It does matter when two points are
brought together.  The most plausible form of
$L_D$ is $L_D^{(1)} = \alpha R_g^{2 \over 3}L_R^{({2 \over 3})}$ via (2)
where $R_g^{2 \over 3} = (\rho^2_i)^{1 \over
3}$ accounts for the time scale implicit in $L_0$.  Alternatively, $R_g^{2
\over 3}$ just gives to $L_D$ the same
scaling dimension as $L_0$; there are no intrinsic scales in the inertial
range.  We will do perturbation theory in
the parameter $\alpha$, the structure of which convinces us that it is
equally plausible to use an analytically more
tractable model which induces the same local singularities into the
eigenfunctions $\psi$, $L_D^{(2)} = \alpha R_g^{2
\over 3}
\beta(\rho)
\nabla^2_\rho$ where $\beta(\rho)$ is homogeneous of degree $4 \over 3$,
inherits the permutation symmetry induced by
$r_i \leftrightarrow r_j$ on the $\rho$'s, and $\beta(\rho) \sim r_{ij}^{4
\over 3}$ as any pair
$(i, j)$ coalesce; a ``Kolmogorov point".  Near such a point $\psi$ has
local s-wave symmetry in $\vec r_{ij}$.  It
will also be useful to consider a scale invariant Laplacian model
$L^{(0)}_D = {({\alpha \over {2 n}})}R_g^2
\nabla^2_\rho$.

For $\alpha \ll 1$, $L_D$ is a singular perturbation to $L_0$, in
particular near a Kolmogorov point we can
chose the $\vec \rho_i$ such that $\rho_1 \rightarrow 0$ and $L_D \sim
\rho_1^{4 \over 3} \nabla^2_{\rho_1}$
whereas $L_0$ involves powers of $\rho^a_1 \partial^b_{\rho_1}$.  Now when
$\rho_1 \rightarrow 0$ the
homogeneous Hopf equation becomes,

$$L^{({n -1})}_0 <\theta^2(1) \theta(2)..> + \alpha R^{2 \over 3}_g
<\epsilon(1) \theta(2)..> = 0\eqno(4)$$

\ni
(plus $O(\alpha)$ less singular terms in the case of $L^{(1)}_D$).  The new
operator $\epsilon$ is the
local dissipation rate, expressible as  $L^{({2 \over 3})}_R (\rho)
\theta(\rho +1) \theta(1)$ in the inertial
range and matching to $\kappa(\nabla\theta)^2$ in the dissipation range.
Of course (4) is not a closed
equation, but $<\epsilon\theta ..>$ must have a positive scaling dimension,
(which can be demonstrated by observing from (1) that $\displaystyle{\sup_{r}}
|\theta(r)|$ and $<\epsilon>$ are finite and set by the large scales).
Hence, the exponent for $<\theta\theta ...>$
must exceed ${2 \over 3}$.$^{11}$  Scaling and the generic behavior of the
correlator when any two points coalesce, insure that $\epsilon$ carries a
dimension of ${2 \over 3}$ relative to $\theta^2$ in all correlators.

To proceed with an explicit calculation, the physical coordinates
$\rho$ can be factored, e.g. in the case $(n, d) = (2, 3)$: $\rho^a_i
= \sum_i\prime R_{ii\prime}(\chi) \ \xi_{i\prime} \
\eta^a_{i\prime}$
where $R$ represents isospin rotations by $\chi$ and $\vec \eta_{1, 2}$ are
two orthogonal unit vectors
defining a representation for $SO(3)$; $(\vec \eta_3 \equiv \vec \eta_1
\wedge \vec \eta_2)$. Suitable, ``gauge"
conditions must be imposed on functions defined on the $\chi$, $\xi$,
$\eta$ manifold if they are to reassemble into
smooth functions of $\rho$.

Let $h_\nu(x_\alpha)$ be a homogeneous function of degree $\nu$ in its
arguments, $\alpha = 1-3$, then any function of the form
$\psi_\circ = \int e^{iq\phi} h_\nu(n_i \rho^a_i e^a_\alpha) D^l_{m,o}
(\hat e)d\phi d\hat e$ is an $n = 2$ eigenvector
of $G^2$ and $L^2$ with eigenvalues $\nu(\nu +1)$ and $l(l + d - 2)$
provided $n_1 + in_2 = e^{i\phi}$
and $D$ is a representation of
$SO(3)$ defined by the triad $\hat e_\alpha$.  We can also multiply
$\psi_\circ$ by any element of $\vec
\rho_1 \wedge \vec \rho_2$ since $G_{ij}\vec \rho_1 \wedge \vec \rho_2 =
0$.  The $\rho$ dependence of all these
functions can be written interms of $\Bigr(w = 2\xi_1 \ \xi_2/(\xi^2_1 +
\xi^2_2)\Bigl)$

$$\psi^\lambda_{\nu , q, l, m, m\prime} = e^{iq\chi}(\xi_1\xi_2)^{\lambda
\over 2} \ P_\nu^{q,
m\prime}(w^{-1})D^l_{m, m\prime}(\eta)\eqno(5)$$
\ni which has a complete set of 6 quantum numbers and defines the sense in
which (3) is
integrable.  An integral representation for the Jacobi function, $P$, was
already given in Eq. (6)
of Ref. (6), and
$\nu$ is related to ${\lambda \over 2}$ by an integer due to boundary
conditions at $w = 0$.  The
index $m^\prime$ corresponds to rotating on the isospin index of $\vec
\eta_{1, 2}$ and is not
summed.

Acting on a general function of $\phi(w, \eta_i)$ with the $\lambda$, and
$\chi$, dependence of $\psi$ as in (5)

$$e^{-iq\chi}(\xi_1\xi_2)^{-\lambda \over 2}G^2 \psi = w^2\partial_w(1 -
w^2)\partial_w\phi + {1 \over 4}
{{2wqI_3} - {({q^2 + I^2_3)w^2}} \over {(1 - w^2)}}\phi\eqno(6a)$$

\ni
where $I_3 = i^{-1}(\eta_1 \cdot \partial_2 - \eta_2 \cdot \partial_1)$.
Reexpressing the Laplacian,

$$\eqalign{e^{-iq\chi}(\xi_1\xi_2)^{-\lambda \over 2} {R^2_g \over 4}
\nabla^2_\xi \psi &= \Bigr[\partial_w (1 -
w^2)\partial_w + (1 + \lambda)(1 - w^2)w^{-1}\partial_w + {1 \over
4}\lambda^2/w^2 +\cr
{1 \over 4} (2qwI_3 - q^2 - I^2_3)/
&(1 - w^2)  + {1 \over 2}w^{-1}\Bigl({\xi_2
\over \xi_1}(\eta_3 \cdot \partial_1)^2 + {\xi_1 \over \xi_2}(\eta_3 \cdot
\partial_2)^2\Bigl)\Bigl]\phi\cr}\eqno(6b)$$

For the skewness, the most relevant eigenvalue of (3) has $l = 1$ so we can
write $\phi = \phi_1(w)\eta_1 + i
\phi_2(w)\eta_2$, then $I_3 \rightarrow \sigma_x$ ($\sigma$ are the Pauli
matrices) acting on
$(\phi_1, \phi_2)$, $(\eta_3 \cdot \partial_a)^2 \ \eta_b =
-\delta_{ab}$, and ${\xi_1 \over \xi_2}$ is a function of w.
It suffices to treat $0
\leq w \leq 1$, negative $w$ are equivalent to a similarity transformation
of the equation by $\sigma_z$.

Thus the Hopf operator with Laplacian damping, $L_0 + L^0_D$, reduces to a
pair of second order
equations, for which the eigenvalue problem can be solved numerically by
shooting from near $w \sim 1$, where
$L_0$ dominates; to $w \sim 0$ or $\rho_{1, 2}$ parallel $(n, b, \vec
\rho_1 \wedge \vec \rho_2 =
\xi_1\xi_2\vec\eta_3)$ which is an invariant set under the coherent part of
the velocity field.  The eddy
damping, i.e., (6b), dominates there and our normalizations require us to
impose that $\phi$ diverge no more
rapidly than $w^{-\lambda \over 2}$.  The most relevant eigenvalue with the
3-fold permutation symmetry
required for $<\theta\theta\theta>$ is $q = \pm 3$, followed by $q = \pm
9$, since there is a gauge requirement
for odd $q$ when $l$ is odd.  We find $\lambda - 1 = {18 \over 5}\alpha -
3.066\alpha^{3 \over 2} + 1.436\alpha^2$, the
first term is analytic (see below) while the others are fit to numerical
data accurate for
$\alpha \leq 1$.  Thus $\lambda$ is nonuniversal, i.e.,
$\alpha$ dependent, but stays close to its anomalous value in experiments
of $\sim 1$, all the more so since the correction term need not be positive
($\lambda_3$ need not be greater than 1); e.g., for certain models of K41
eddy damping. The
eigenvector can be thought of a superposition of modes of the form (5) with
respect to the discrete allowed $\nu$ at
fixed $\lambda$.$^6$

We now set up perturbation theory via matched asymptotic expansions with
sufficient generality that it
applies to the $L_D^{(1)}$ case with $\beta = R^{4 \over 3}_g \ \buildrel \sim
\over \beta/4$,
$\buildrel \sim \over \beta = \beta_0(\chi)\beta_0(\chi - {\pi \over 3})
\beta_0(\chi + {\pi \over 3})$,
$\beta_0 = 2^{-{2 \over 3}}(1 - {\rm cos}(2\chi)\sqrt{1 - w^2})^{2 \over
3}$.  The Kolmogorov points
correspond to $w=0$ and $\chi = 0$, $\pm {{ 2 \pi } \over 3}$,
and we have made the convention $|\xi_1| \leq \xi_2$.
The transformation $\chi \rightarrow \chi \pm {{ \pi }
\over 3}$ corresponds to a permutation and reflection of $r_i$
implying that $\phi$ is antiperiodic in $\chi$
with period $ {\pi\over 3}$ and of course periodic with period
${{2\pi} \over 3}$.  Also $\phi$ is even under interchange of $r_{1, 2}$
i.e., $\rho_1 \rightarrow - \rho_1 \
\rho_2 \rightarrow \rho_2$  or $(\chi,\xi_1) \rightarrow -(\chi,\xi_1)$,
$\xi_2,\vec \eta_1$ unchanged.  For $w=0$, $\phi$ is even
around $\chi = 0$, which together with
its antiperiodicity implies $\phi(\chi = \pm {\pi \over 6}) = 0$.

To go through the region $w^2 \sim \alpha$ where $L_0$ and $L_D$ are
comparable, we have to solve the
following crossover equation, $(\buildrel \sim \over \phi \equiv
w^{-{\lambda \over 2}}
\phi)$

$$\eqalign{\Bigl(w^2\partial^2_w - w\partial_w &+ {1 \over 4}
w^2\partial^2_\chi - {1 \over 2}
iw\sigma_x \partial_\chi\Bigl)\buildrel \sim \over \phi +
\alpha\beta\Bigl(\partial^2_w + {1
\over w}\partial_w
+ {1 \over 4}\partial^2_\chi + {5 \over 4} - {1 \over 4} (1 - \sigma_z)\cr
&- {1 \over 2} (1 + \sigma_z)/w^2\Bigl)\buildrel \sim \over \phi + {5 \over
3}(\lambda - 1)\buildrel \sim \over \phi
= 0\cr}\eqno (7)$$

\ni
where we have replaced $q^2$ by $-\partial^2_\chi$.  Near the Kolmogorov
points generally $\buildrel \sim \over \beta
\sim (w^2 + 4
\chi^2)^{2 \over 3}$ and the $O(\alpha)$ derivative terms in (7)
will be recognized as a 3-D Laplacian in
cylindrical coordinates $(w, \chi)$.  Elsewhere
$\partial_\chi \ll \partial_w$.  For  $\alpha \ll w^2 \leq 1$ the solution
of (7) must reduce to a linear
combination of solutions (5) to (3) with $\nu =1/2$ and scaled with
$R_g$
which for small $w$ behave as $\phi^0_q = \Bigr(
{1 \over 2}|q|w + {1 \over 4}(1 - q^2) w^2,
(1 - {1 \over 8}q^2w^2){\rm sgn}(q) \Bigl)/(q^2 - 1)$.
This has to be propagated via (7)
to $0 \leq w^2 \ll \alpha$ which is
particularly simple for $\phi_2$ which is found to behave as $(a(\chi) + {1
\over 8}\partial^2_\chi a \ w^2 ...)$. Substituting into (7)
we derive the eigenvalue problem along $w = 0$ as

$$\alpha\buildrel \sim \over \beta(\chi) (\partial^2_\chi + 1)a + {20 \over
9}(\lambda - 1)a = 0\eqno(8)$$

\ni
The eigenvalue for Laplacian dissipation follows by setting $\buildrel \sim
\over \beta = 1$ and $a = e^{iq\chi}$ while
in the Kolmogorov case we integrate from $\chi = 0$ to the antiperiodic
point $\chi = {\pi \over 3}$ with boundary
conditions $a \sim \Bigl(1 + O(1 - \lambda)|\chi|^{2 \over 3} + O(\chi^{4
\over 3})\Bigr)$ and find
$(\lambda - 1)/\alpha = -0.0707$ (versus $-0.2$ for d=2).  The eigenfunction is
monotonic and largely
fixed by the symmetries and the
boundary conditions at $\chi = 0$, ${\pi \over 3}$.

Neither the computed $a(\chi)$ nor $\lambda$ can be directly compared
with the experiment since they depend
on details of the eddy damping.
However we can make testable predictions by
expressing $\psi = <\theta_1 \theta_2 \theta_3>$ for $\rho_i$ parallel
($\xi_1 = 0$ by convention  and $tan(\chi) = \rho_1/\rho_2$)
as $\xi^\lambda_2 a(\chi) \vec \eta_2 \cdot \vec g$ to within an overall
constant.   Then away from colinearity ($w^2 > \alpha$),
the solution is a superposition of "SL(n) modes"
(5) weighted by $a_q$, the Fourier transform of $a(\chi)$.
More explicitely we use the integral
representation of $\psi$ and write a convolution

$$\psi(\rho) = \vec g \cdot \int^\pi_0{{({\rm cos}(\phi)\xi_1 \vec \eta_1 +
{\rm \sin}(\phi)\xi_2\vec \eta_2)}
\over ({\rm cos}^2(\phi)\xi^2_1 + {\rm \sin}^2(\phi)\xi^2_2)^{{1 \over 2} -
{\lambda \over 2}}} h(\phi -
\chi)d\phi\eqno(9)$$

\ni
where $h_q = {\rm sgn}(q) (q^2 - 1) a_q$ defines the
$q$ values in (5) that are
superimposed, $\nu = {\lambda \over 2}$, and we summed on $m\prime$ to
impose the symmetry under $r_{1, 2}$
interchange.  To be consistent, we
should expand the denominator
to first order in $(\lambda - 1)$ (the $0^{\rm th}$ order term vanishes).
Then for $w \ll 1$, $\phi^0_q$ is just the transform of the
($\xi,\eta$) factor in (9) which explains the relation of $h$ to $a$.

The flatness in two spacial dimensions is illustrative of correlation
functions with $n > d$.  By interchanging the
meaning of space and isospin labels we can use the same $\chi$, $\xi$,
$\eta$ coordinates as before but to emphasize
their new meaning we put quantum numbers $p, q, q\prime$ on $D$ in (5) and
replace $q$ by $m \rightarrow 0$ for the
s-wave state.  Equations ($6a, b$) also remain valid though it is now
convenient to recognize that the
$\eta_3 \cdot \partial_a$ (a=1,2) together with $I_3$ comprise a spin algebra
operating on the $q\prime$ index of
$D^p_{qq\prime}(\eta)$.

For $n = 2$, the isospin was Abelian, and for Laplacian damping the most
relevant eigenmode had $q = 3$.  The quantum
numbers for the analogous flatness mode follow by writing
$\displaystyle{\prod^{4}_{1}} g \cdot r_i$ interms of
$\rho_i$ which gives a particular combination of $(q, p) = (0, 4)$ and (3,
4).  Note only $p$ and $q\prime$ appear in (6b) not
$q$.  Thus the solution $\sim \displaystyle{\sum_{q\prime}}
\phi^{q\prime}(w)D^p_{qq\prime}(\eta)$ to the Hopf
equation $L_0 + L_D^{(0)} = 0$ can be sought in a $2p + 1 = 9$ dimensional
space labeled by $q\prime$ which further
symmetries reduce to a three dimensional sub space $(0, 2, 4)$ even under
$q\prime \leftrightarrow - q\prime$.  The
eigenvalue is $\lambda_F = {45 \over 16} \sqrt {2\alpha /3} +2.2 \alpha - 3.8
\alpha^{3/2} + 1.64 \alpha^2...$. As before the leading term can be obtained
analytically by solving the crossover equation.

The analytic details of the perturbative theory for $\lambda_F$ and its
eigenmode for either $L_D^{(0)}$ or
$L_D^{(2)}$ damping resemble the $n = 2$ case in that a solution to (3) for
$\alpha \ll w^2 \leq 1$ has to be matched
to one for $0 \buildrel < \over \sim w^2 \ll \alpha$.  The geometry differs
because before the Kolmogorov points, e.g.
$|\rho_1| = 0$ or $\chi = w = 0$, were subsets of the condition for
parallism, $w = 0$, whereas for the $2D$ flatness the Kolmogorov points
are for instance $\eta^a_1 = 0$, $(a = 1,
2$ is space), while the condition all three $\vec \rho_i$ parallel is still
$w = 0$.  The two sets have only a lower
dimensional intersection.  The eigenvalue problem analogous to (8), still
operates within $w = 0$ but the space is
$SO(3)$.  The integral representation replacing (9) will apply for $w^2 \gg
\alpha$ and also away from the Kolmogorov
points.

It is noteworthy that the dissipation is a bigger perturbation on the
flatness than the skewness i.e., $\sqrt \alpha$
vs $\alpha$; but we have not seen how $\lambda_F \geq {2 \over 3}$ emerges
from the calculation, so for the moment we
take this as a lower bound on $\alpha$.

The symmetry breaking external gradient $\vec g$ is a relevant pertubation
in that normalized
high order odd moments of $\partial\theta$, which vanish under isotropic
conditions, will grow
as $R^{{3 \mu_N \over 4}}$ when $\mu_N = N/3 - \lambda_N < 0$
where $\lambda_N$ is the N point $\theta$ correlator.
Thus the derivative skewness
may already diverge ($\lambda_S < 1$ for the pseudo Kolmogorov
dissipation $L_D^{(2)}$ ), and for
small $\alpha$ all higher odd moments certainly diverge,
since $\lambda_N = 1 + O(\alpha)$.$^{12}$
Note for even $N$, $\mu_N$ is also the scaling exponent for the $N/2$ point
dissipation correlator, i.e., $<\epsilon(1) \epsilon(2)..> \sim R_g^{-\mu_N}$.

Finally, a symmetry
based classification also leads one to ask qualitative questions such as
whether for physical values of the
dissipation the nominally subleading $p$-wave exponent for flatness and other
even moments could be more relevant than the
$s$-wave exponent  and whether there could be a scaling crossover as
the points depart from collinearity.

Our phenomenological approach is based on an approximate symmetry and its
experimental consequences, e.g. (9), should
be more general than the specific equation we solved. We hope also that
the $SL$ invariance is a more robust property of the dynamics
than the scaling exponents and therefore visible at lower
Reynolds' numbers. Our dissipation was adhoc, and for
$n \geq 3$ should look more
hierarchical.  Yet compared to the study of the exact Hopf equation for the
white noise$^{5,7,-10}$ exploring different dissipation
models has the advantage of confronting one with the non-uniqueness and
the non-universality of the problem at hand.  In particular,
exponents are non-universal.$^{13}$

This conclusion may be regarded as
a peculiarity of the passive scalar problem defined here for arbitrary
random velocity fields, in contrast to the properly turbulent velocity
field governed by the Navier-Stokes
equation which for large R, according to K41, posseses
universal statistical properties. Yet, our analysis illustrates
how the large scale anisotropy can affect the small scale statistics.
Applied by analogy to the velocity field, it would suggest the possibility
that the small scale velocity statistics may in fact depend on the
large scale boundary conditions leading to
non-universal exponents for sufficiently high order
correlators.$^{14}$  This of course would just be a manifestation of the
perhaps not so rare "rare" events directly coupling the large and the
small scale of the flow.

EDS thanks the NSF for support under grant DMR 9121654, and both authors
acknowledge the Aspen Center for Physics where this work was initiated.
\vej

\ni{\ub {References}}
\vskip 10pt

\mm{1)}  A. S. Monin and A. M. Yaglom, Statistical Fluid Mechanics, MIT
press, (1975).
\mm{2)}  A. Antonia, E. J. Hopfinger, Y. Gagne, and F. Anselmet, Phys.
Rev. {\bf A30}, 2704 (1984), D. R. Dowling and P. E. Dimotakis, J. Fluids
Mech. {\bf 218}, 109
(1990).
\mm{3)}  P. G. Mestayer, J. Fluid Mech. {\bf 125}, 475 (1982), K. R.
Sreenivasan, Proc. Roy. Soc. Lond., {\bf A434}
165 (1991).
\mm{4)}  M. Holzer and E. Siggia, Phys. Fluids, {\bf 6}, 1820 (1994),  A.
Pumir, Phys. Fluids, {\bf 6}, 2118 (1994),
C. Tong and Warhaft, Z., Phys. Fluids {\bf 6}, 1820 (1994).
\mm{5)}  R. H. Kraichnan, Phys. Rev. Lett., {\bf 72}, 1016 (1994), S. Chen
and R. H. Kraichnan private communication.
\mm{6)}  B. I. Shraiman and E. D. Siggia, C.R. Acad. Sci. Paris {\bf 321},
279 (1995).
\mm{7)}  M. Chertkov, G. Falkovich, I. Kolokolov, and V. Lebedev, Phys.
Rev. {\bf E 51}, 5609 (1995).
\mm{8)}  K. Gawedzki and A. Kupiainen, Phys. Rev. Lett. {\bf 75}, 3834 (1995).
\mm{9)}  B. I. Shraiman and E. D. Siggia, Phys. Rev. {\bf E 49}, 2912 (1994).
\mm{10)} R. H. Kraichnan, Phys. Fluids {\bf 11}, 945 (1968).
\mm{11)} In order to match the inertial range Hopf solutions to those
in the dissipation range all the zero modes are required.
In particular those with
negative $\lambda = -|\lambda_d|$
then appear in the inertial range as terms scaling as
$(\eta/R_g)^{|\lambda_d|}$ where $\eta$ is the
Kolmogorov scale; down by factors of $R$ but capable of influencing
inertial range experiments if
$|\lambda_d| \ll 1$.
\mm{12)} The perturbation theory for Laplacian damping and $d=2$ has been
done for general N: $\lambda_N = (-1.13 + 1.27 N) \sqrt {\alpha/(N-1)}$
for N even and $\lambda_N - 1 = 1.89 N \alpha/(1+6.00 N^{-2} + 95.35 N^{-4} )$
for N odd.  In both cases the expressions are fits of a complicated
analytic formula good to 1\%.
\mm{13)} Pertubation around the white noise limit has been shown to give
nonuniversial results
by M. Chertkov, G. Falkovich, and V. Lebedev (preprint).
\mm{14)} A. Pumir and B. I. Shraiman, Phys. Rev. Lett. {\bf 75}, 3114 (1995).
\end